\documentclass[aps,pre,twocolumn,amsmath,nofootinbib,10pt,floatfix]{revtex4-2}

\usepackage[english]{babel}
\usepackage{graphicx}
\usepackage{ulem}
\usepackage{soul}
\usepackage{float}
\usepackage{hyperref}
\hypersetup{colorlinks,linkcolor=red,citecolor=blue,breaklinks=true}
\usepackage{amsfonts}
\usepackage{bm}
\usepackage{bbm}
\usepackage{verbatim}
\usepackage{esint}
\usepackage{marginnote}
\usepackage{feynmf}
\usepackage{xcolor}
\usepackage{cleveref}
\usepackage{chngcntr}

\renewcommand{\vec}[1]{\boldsymbol{#1}}
\newcommand{\RNum}[1]{\uppercase\expandafter{\romannumeral #1\relax}}

\def \beq {\begin{eqnarray}}
\def \eeq {\end{eqnarray}}

\def \pcinfty {0.645}
\def \pcinftyerr {0.002}
\def \nue {1.3}
\def \nuerr {0.2}
\def \fiso {2.2}
\def \fisoerr {0.3}
\def \zetae {0.25}
\def \zetaerr {0.1}
\def \faniso {4.0}
\def \fanisoerr {1.0}

\def \pcinftycollapse {0.646}
\def \nucollapse {1.3}
\def \fisocollapse {2.2}

\def \figwid {1.0}

\begin{document}

\title{Rigidity transitions in anisotropic networks happen
	 in multiple steps}
\author{William Y. Wang$^1$, Stephen J. Thornton$^1$, Bulbul Chakraborty$^2$, Anna Barth$^1$, Navneet Singh$^1$, Japheth Omonira$^1$, Jonathan A. Michel$^3$, Moumita Das$^3$, James P. Sethna$^1$, Itai Cohen$^{1,4}$}

\affiliation{
$^1$Department of Physics, Cornell University, Ithaca, New York 14853, \\
$^2$Department of Physics, Brandeis University, Waltham, Massachusetts 02454, \\
$^3$School of Physics and Astronomy, Rochester Institute of Technology, Rochester, New York 14623, \\
$^4$Department of Design Technology, Cornell University, Ithaca, New York 14853} 
\date\today

\begin{abstract}
    We study how the rigidity transition in a triangular lattice changes as a function of anisotropy by preferentially filling bonds on the lattice in one direction. We discover that the onset of rigidity in anisotropic spring networks arises in at least two steps, reminiscent of the two-step melting transition in two dimensional crystals. In particular, our simulations demonstrate that the percolation of stress-supporting bonds happens at different critical volume fractions along different directions. By examining each independent component of the elasticity tensor, we determine universal exponents and develop universal scaling functions to analyze isotropic rigidity percolation as a multicritical point. We expect that these results will be important for elucidating the underlying mechanical phase transitions governing the properties of biological materials ranging from the cytoskeletons of cells to the extracellular networks of tissues such as tendon where the networks are often preferentially aligned.
    
\end{abstract}

\maketitle

\section{Introduction}
Rigidity percolation in central-force lattice models has emerged as an important tool for modeling structural networks in cells and cellular tissues
\cite{AnisotropicCollagen,AnisotropicSpringNetworks,wyse2022structural,silverberg2014structure}. Such central-force lattices consist of harmonic springs connecting nodes. The network is randomly filled by introducing springs between nodes to achieve a density $p$, which denotes the fraction of occupied bonds in the network. At low bond occupation, the bond network does not span the entire system. As $p$ increases, the network undergoes a percolation transition where a cluster of bonds can now span the entire network. This tenuous cluster can only support stresses if there are angular forces between bonds~\cite{RedundancyCrossLink}. In many practical scenarios, such bond bending forces are small compared with bond stretching. In such cases, the contribution to rigidity from bond bending is ignored. In this scenario, the network remains floppy until $p$ reaches the so-called rigidity percolation threshold where bond stretching is activated under infinitesimal deformation of the network.

Rigidity percolation has been well studied in isotropic networks under different bending and stretching constraints \cite{CriticalityIsostaticity}. The onset of this transition, however, is sensitive to details of the bond distributions. For example, previous work has shown that including structural correlation within isotropic networks can result in significant changes in the critical bond occupation threshold for rigidity percolation \cite{CorrelatedGels, ReentrantCorrelated}. Furthermore, studies have shown that straining a percolated but floppy network, such as by shearing it in one direction, can drive a rigidity transition \cite{FredStrain}. For example, straining the network preferentially along the maximum extension axis activates bond stretching, which rigidifies the network. Finally, previous computational studies have also modeled anisotropic networks through an anisotropically diluted triangular lattice and found that the onset of rigidity agrees well with Maxwell constraint counting and that the system can be approximated using an effective medium theory (see \cite{AnisotropicSpringNetworks} and Appendix~\ref{EMTApp}). Missing from these analyses, however, are detailed investigations of whether the critical exponents and scaling functions characterizing the rigidity transitions depend on these details of the bond distributions. Measuring the critical exponents and how they depend on the bond distributions is critical for determining whether such mechanical phase transitions are in the same universality class, which informs the relevant physics governing these transitions. Understanding this physics is an important step for learning how to control these transitions in materials ranging from biological tissues to synthetic fiber networks.   

Here, we will examine the critical exponents and locations of phase boundaries in anisotropic networks. Though our model is quite similar to ones previously investigated \cite{AnisotropicSpringNetworks}, we find slightly different critical exponents for the isotropic rigidity percolation transition. Remarkably, we also discover that when we tune away from isotropy, the network exhibits \textit{two rigidity transitions}. The first is associated with the component of the strain tensor for stretching along the preferential filling direction. The second rigidity transition is associated with the remaining components of the strain tensor. As such we find that the isotropic rigidity transition is a multi-critical point from which anisotropic rigidity transitions emanate (Fig.~\ref{fig:PhaseDiagram}).

\section{Methods}
\subsection{Model for anisotropic rigidity percolation}

We generate triangular lattices (coordination number $z=6$) of $L^2$ sites and periodic boundary conditions in both directions. Bonds are diluted based on their orientation, where $p$ denotes the fraction of occupied bonds in the network. Anisotropy is introduced during lattice generation by filling bonds preferentially based on their orientation. We define the ratio $r$ as the probability of bond occupation along the horizontal direction divided by the probability of bond occupation in the other two independent directions. We then build lattices using methods similar to those previously developed (see Appendix~\ref{BondFilling} for details) and investigate the regime where $r\geq1$. Importantly, we are able to adjust $r$ without changing $p$, which allows us to shuffle bonds and investigate how long-wavelength anisotropy affects the scaling of moduli in equally dense networks. In these coordinates, $r=1$ represents a completely randomly diluted triangular lattice and $r=\infty$ a lattice which has bonds only in the horizontal direction. Our choice of anisotropy for generic $r$ results in four independent long-wavelength components of the elasticity tensor, $\{C_{xxxx},\ C_{yyyy},\ C_{xyxy},\ C_{xxyy}\}$, each of which can be extracted for the various lattices.

\subsection{Simulation details}
We measure the components of the elasticity tensor for each random lattice realization at different values of filling fraction $p$, anisotropy $r$, and linear system size $L$. To measure these components, we first apply a small external strain $\varepsilon_{ij}$ of magnitude $10^{-3}$. In the regime of linear elasticity, the energetic cost of such a deformation is quadratic in the strain:
\beq
E=\frac{1}{2}\varepsilon_{ij}C_{ijkl}\varepsilon_{kl}.
\label{eqn:elasticTensor}
\eeq
We apply strains $\varepsilon_{xx}$, $\varepsilon_{yy}$, and $\varepsilon_{xy}$ to measure the elastic coefficients $C_{xxxx}$, $C_{yyyy}$, and $C_{xyxy}$ directly. To measure $C_{xxyy}$, we perform a bulk compression and subtract out the energetic contributions from the independently measured $C_{xxxx}$ and $C_{yyyy}$ moduli.

In order to measure the energetic costs of our imposed strains in the disordered lattices, we minimize the central-force energy functional over the positions of the nodes. To capture the linear response, we truncate to leading order in the displacement of vertices:
\beq
    E = \frac{1}{2}\sum_{\left\langle ij\right\rangle}k_{ij}\left(\mathbf{u}_{ij}\cdot\hat{\mathbf{r}}_{ij}\right)^2
\label{eqn:linearizedEnergy}
\eeq
where $\mathbf{u}_{ij}$ is the difference
between the displacement vectors for vertices $i$ and $j$, and $\hat{\mathbf{r}}_{ij}$ is defined as the unit vector between vertices $i$ and
$j$ in the initial configuration. The spring constant $k_{ij}$ connecting sites $i$ and $j$ is either 0 or 1, according to the random number seed, $p$, and $r$. For each type of imposed fixed-amplitude strain, we minimize Equation~(\ref{eqn:linearizedEnergy}) and use the resulting energy and Equation~(\ref{eqn:elasticTensor}) to extract the values of the independent moduli (see Appendix~\ref{AppNumerics} for details).

We then individually perform scaling analyses for each independent component of the elasticity tensor. Starting with a rigid network, we repeatedly remove bonds and minimize the energy for each strain until the network becomes sufficiently ``floppy'' in all directions; here, a network is considered floppy in a particular direction if the corresponding modulus falls below a threshold of $G_\text{min} \equiv 10^{-8}$, which is the simulation tolerance. The moduli as a function of $p$ were averaged for each system size and orientation strength pair ($L, r$), sampled over $10^2 \mbox{--} 10^4$ random seeds. This procedure allows us to perform a scaling analysis of each component of the elasticity tensor separately.  
\section{Results}
\subsection{Isotropic Networks}

We begin by focusing on the long-wavelength isotropic case, where bonds are removed without regard to their orientation ($r=1$). We determine the value of $p$ where each lattice becomes able to support a stress, defined to be $p_c$. To extrapolate our results to infinite lattices, we conduct a finite-size scaling analysis (Appendix~\ref{AppDistCollapse}). For a given system size, we find the rigidity threshold for many different lattices and create a histogram of these $p_c$ values. We find that for increasing lattice sizes, $L$, the width of the histogram for the threshold values $p_c$ decreases as $L^{-1/\nu}$. We also find that the mean value of the histogram, $\langle p_c\rangle_L$, approaches a value $p_c^{\infty}$, the threshold in the infinite system, with the same power law: 
\beq
    \langle p_c \rangle_L -  p_c^{\infty} \sim L^{-1/\nu}.
\eeq

\begin{figure}[!ht]
\begin{center}
    \includegraphics[alt={phase diagram}, width=\figwid\linewidth]{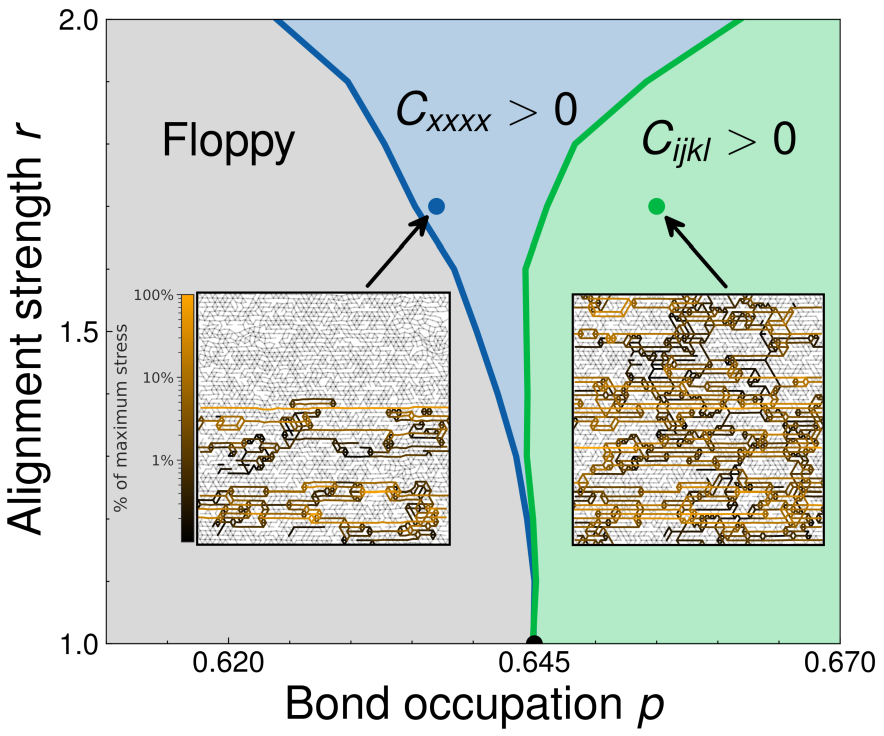}
\end{center}
\caption{\textbf{Rigidity percolation phase diagram and rigid clusters.} The location of the phase boundaries are found in the thermodynamic limit using a finite-size scaling analysis. The insets show an anisotropic network close to its rigidity percolation point for the $C_{xxxx}$ modulus (left) and $C_{ijkl} \neq C_{xxxx}$ moduli (right). The shading corresponds to the energy contributed by each bond after a strain in the $x$ direction (gray indicating no stress and brighter indicating higher stress).}
\label{fig:PhaseDiagram}
\end{figure}

Our analysis determines that $p_c^{\infty}=\pcinfty \pm \pcinftyerr$, depicted by the black dot along the horizontal axis of Fig.~\ref{fig:PhaseDiagram}, and $\nu = \nue \pm \nuerr$. The location of the threshold at $p_c^{\infty}$ indicates a small deviation from the Maxwell counting constraint, which states that the 2D triangular lattice to have $2$ constraints per site, $p=2/3$ of the lattice must be occupied. The deviation from this prediction in our measured value of $p_c^{\infty}$ is similar to what is found in other works~\cite{CriticalityIsostaticity}. Importantly, we find that in the isotropic system ($r=1$) all the moduli share the same threshold value $p_c^{\infty}$.

\begin{figure}[!ht]
\begin{center}
\includegraphics[alt={isotropic scaling function}, width=\figwid\linewidth]{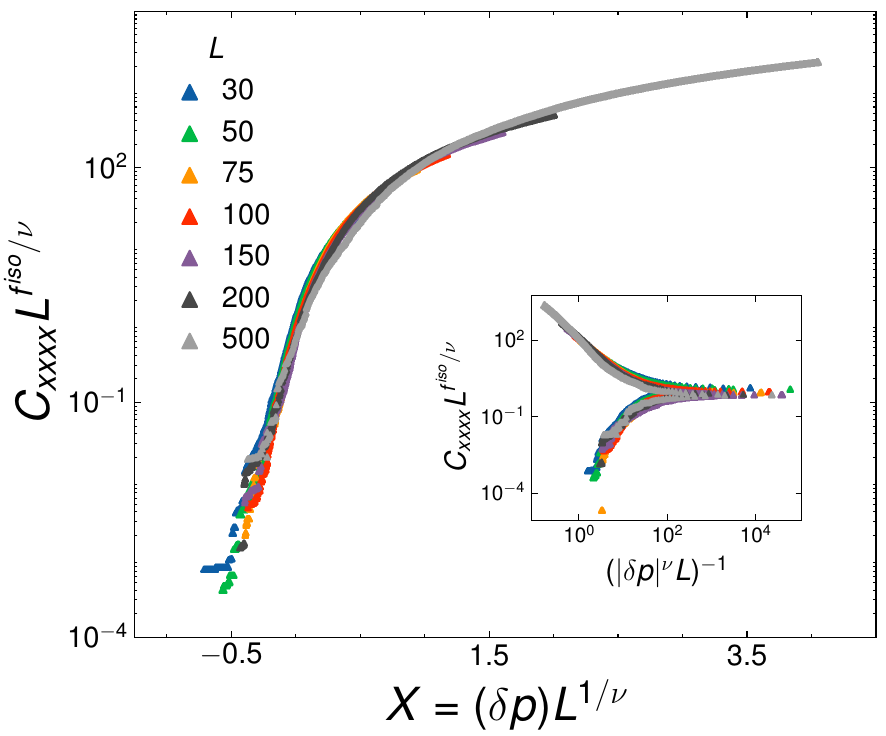}
\end{center}
\caption{\textbf{Universal scaling function for the $C_{xxxx}$ component of the elasticity tensor at isotropy $(r=1)$.} All data for this component of the elasticity tensor collapse onto a single curve $\mathcal{C}_{xxxx}^{\textrm{iso}}$ when plotted against the finite-size scaling variable $X \equiv (\delta p)  L^{1 / \nu}$. The inset shows the same collapse against the scaling variable $(|\delta p|^\nu L)^{-1}$ (on a log-log scale). See Appendix~\ref{AppIsoModulusCollapse} for a similar analysis of the other components of the elasticity tensor. We use $p_c^\infty = \pcinftycollapse$, $\nu = \nucollapse$, and $f^{\textrm{iso}} = \fisocollapse$ to obtain excellent collapse for all the components of the elasticity tensor. 
}
\label{fig:UniversalIso}
\end{figure}

Next, we perform a finite-size scaling analysis of each component of the elasticity tensor, which admits a scaling
\beq
    \begin{aligned}
        C_{ijkl}\left(p, L \right) &= L^{-f^{\textrm{iso}}_{ijkl} / \nu} \mathcal{C}_{ijkl}^{\textrm{iso}}\left( X \right) \\
        X &\equiv (\delta p)  L^{1 / \nu},
    \end{aligned}
    \label{eqn:IsotropicModulusScaling}
\eeq
where $\delta p \equiv p - p_c^\infty$. Thus, in principle each modulus component could have a different scaling exponent, $f^{\textrm{iso}}_{ijkl}$, and shape, $\mathcal{C}_{ijkl}^{\textrm{iso}}$. The universal function $\mathcal{C}_{xxxx}^{\textrm{iso}}\left(X\right)$ is plotted along with the appropriately rescaled data for $C_{xxxx}$ in Fig.~\ref{fig:UniversalIso}. We find excellent scaling for the different system sizes. In the inset of Fig.~\ref{fig:UniversalIso}, we plot the collapsed data against the equivalent scaling variable $\left|X\right|^{-\nu}$, which is a more common choice of variables in the literature, but leads to two branches of the scaling function. We find similarly excellent collapses for all the modulus data (see Fig.~\ref{fig:AllModuliCollapse} for the collapse of other components of the elasticity tensor). For all of these analyses we use the threshold $p_c^\infty=\pcinftycollapse$, the critical exponent $\nu=1.3$, and obtain $f^{\textrm{iso}}_{ijkl}=f^{\textrm{iso}}=\fiso \pm \fisoerr$. Thus, every independent component of the linear elasticity tensor appears to vanish as $\left|\delta p\right|^{f^{\textrm{iso}}}$ (see Appendix~\ref{AppIsoModulusCollapse}).

\subsection{Anisotropic Networks}

We extend our analysis in the previous section to lattices with long-wavelength anisotropy ($r>1$), where we preferentially fill bonds in the $x$ direction. We begin with the determination of the phase boundary. For each value of $r$, we perform a finite-size scaling analysis similar to the one we conducted for the isotropic case: we create histograms of the values of $p$ where $C_{ijkl}$ vanishes, $p_c^{ijkl}(r)$, and determine how their mean values extrapolate to the infinite system. For each value of $r$ we find that the mean location of the critical point is well described by:
\beq
    \langle p_c^{ijkl}(r) \rangle_L  - p_c^{\infty, ijkl}(r) \sim L^{-1/\nu'}.
\eeq

Remarkably, we find that for $r>1$, the transition for the $C_{xxxx}$ modulus is distinct from the threshold values for the other components of the elasticity tensor. We find that the phase boundary for $C_{xxxx}$ bends towards lower values of $p$ with increasing $r$ (Fig.~\ref{fig:PhaseDiagram}). The transition curves for the other moduli appear nearly identical for each system size and bend towards higher values of $p$ with increasing $r$. This separation indicates a region for which the network is only rigid when strained along the preferred bond orientation direction.  We verify these phase boundaries are the same at isotropy but distinct for $r>1$ by measuring the separations between the histograms of $p_c$ values for different components of the elasticity tensor as a function of system size (see Appendix~\ref{AppPhaseSeparation}). Thus, the transition from a floppy to a rigid phase occurs in two stages when the system is anisotropic.

Next, we test the scaling behavior of the elasticity tensor components near the critical points. In principle, including anisotropy could introduce corrections to scaling that bend the phase boundary in a trivial way, leaving all critical exponents the same as in the isotropic system. We tested this scenario, by fixing $r>1$ and attempting to collapse the data near the relevant $p_c(r)$ values for each component of the elasticity tensor, keeping $f^{\textrm{aniso}} = f^{\textrm{iso}}$ and $\nu'=\nu$, but found very poor collapse. This poor collapse suggests that the anisotropic phase transition is in an entirely different universality class, with different values for the critical exponents. We thus conjecture that the vicinity of $r=1$ should be analyzed as a crossover scaling between two distinct critical points.

Inspired by renormalization group approaches, we analyze our data using crossover scaling functions expected to be valid in the vicinity of the isotropic critical point:
\beq
    \begin{aligned}
        &C_{ijkl}\left(p,r,L\right) = L^{-f^{\textrm{iso}} / \nu} \mathcal{C}_{ijkl} \left(X, Y \right) \\
        &X \equiv (\delta p)  L^{1 / \nu},\ Y \equiv (r-r_c)L^{\zeta / \nu}
    \end{aligned}
    \label{eqn:CrossoverScaling}
\eeq
with $r_c\equiv 1$. Note that $\mathcal{C}_{ijkl}(X, 0)$ is equal to the previously defined $\mathcal{C}_{ijkl}^{\textrm{iso}}(X)$ in Equation~(\ref{eqn:IsotropicModulusScaling}). Based on the form of this crossover scaling function, we expect that the moduli depend upon the anisotropy $r$ only through a second scaling variable, $Y$; that is, we expect a scaling collapse of all of our data when plotted against $X$ and $Y$ with a single undetermined exponent $\zeta>0$.

We first estimate $\zeta$ by examining the shape of the phase boundaries in the infinite system away from isotropy. Specifically, as shown in Appendix~\ref{AppAnisoCollapse}, because the arguments of the universal scaling function are invariant scaling combinations, this phase boundary must occur at a fixed value of $X/Y^{1/\zeta}=\delta p/\left(r-r_c\right)^{1/\zeta}$ (with corrections to scaling), so that the separation between the two phase boundaries in Fig.~\ref{fig:PhaseDiagram} scales as $\left(r-r_c\right)^{1/\zeta}$.
From this estimate based on the shape of the phase boundaries, we find $\zeta = \zetae \pm \zetaerr$. 

Many other quantities share this crossover scaling ansatz and allow for independent estimates of $\zeta$. The widths of the histograms of $p_c$ values as we tune away from isotropy are also amenable to a crossover scaling analysis, with the variable $Y$ collapsing these widths (Appendix~\ref{AppAnisoCollapse}). From this scaling collapse, we similarly estimate $\zeta = \zetae \pm \zetaerr$ and find nice collapse (Fig.~\ref{fig:SigmaCollapse}).

\begin{figure}[!ht]
    \centering
    \begin{center}
        \includegraphics[alt={crossover scaling collapse}, width=\figwid\linewidth]{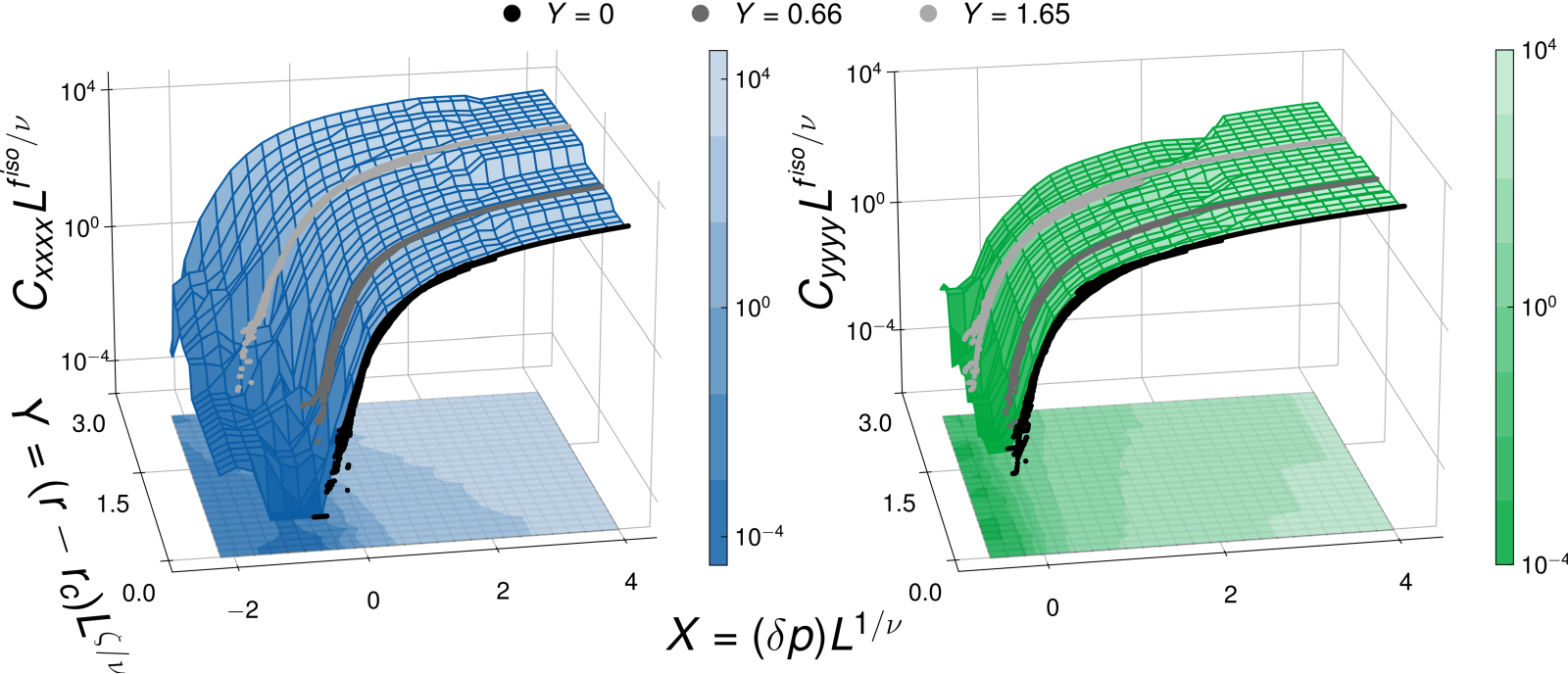}
    \end{center}
    
    \caption{\textbf{Crossover scaling of anisotropic rigidity percolation.} Each independent elastic modulus, a function of the variables $(p,r,L)$, collapses onto a two dimensional sheet when plotted against scaling variables $X\equiv (\delta p) L^{1/\nu}$ and $Y\equiv\left(r-r_c\right)L^{\zeta/\nu}$. (left) The scaling function $\mathcal{C}_{xxxx}\left(X,Y\right)$ and (right) the scaling function $\mathcal{C}_{yyyy}\left(X,Y\right)$. The isotropic data ($Y=0$, Fig.~\ref{fig:UniversalIso}) is scattered in black, and constant values of $Y=0.66$ and $Y=1.65$ are scattered in gray  (Figs.~\ref{fig:CollapseLowY}--\ref{fig:CrossoverHighY}). The height contours are projected onto the $X$-$Y$ plane.}
    \label{fig:SheetAniso}
\end{figure}

This estimate for $\zeta$ and the scaling ansatz in Equation~(\ref{eqn:CrossoverScaling}) can be used to collapse the elasticity tensor components for 250,000 simulations consisting of anisotropy values ranging between $1.0 \leq r \leq 2.0$, bond occupation values ranging between $0.6 \leq p \leq 0.68$, and system sizes $L$ ranging between $30\leq L\leq 500$. We show a two variable scaling collapse for the $C_{xxxx}$ and $C_{yyyy}$ moduli in Fig.~\ref{fig:SheetAniso}. We find excellent collapse of each independent modulus onto a two dimensional sheet. The overlaid data points consist of a portion of the data used to produce the sheet and indicate various slices of constant $Y$: $Y=0$, $Y=0.66$, and $Y=1.65$. The $Y=0$ curve (black) shows the finite-size scaling for isotropic systems and is identical to that shown in Fig.~\ref{fig:UniversalIso}. We observe similarly excellent collapse at the two higher values of $Y$ (Figs.~\ref{fig:CollapseLowY}--\ref{fig:CollapseHighY}). For ease of visualization, we also include height contours projected onto the $X$-$Y$ plane. The height contours for $C_{xxxx}$ curve toward lower values of $X$ as the scaling variable $Y$ increases, reflecting the fact that the corresponding phase transition curves toward lower values of $p$ as $r$ increases. The other moduli, such as $C_{yyyy}$, show the opposite systematic behavior, tending towards higher values of $X$ for increasing $Y$.

The critical exponents determined thus far, including $\zeta$, are properties of the \textit{isotropic} rigidity percolation critical point. We are also interested in the \textit{anisotropic} critical exponents, such as the critical exponent with which each modulus vanishes with $p$ in an infinite anisotropic system ($f_{ijkl}^{\textrm{aniso}}$). The universal scaling function $\mathcal{C}_{ijkl}\left(X,Y\right)$ for the crossover between the isotropic and anisotropic critical points in principle contains all of these anisotropic critical exponents in its singularities in various asymptotic regimes. However, it is quite difficult to get reliable high-precision fits to generic two-variable scaling functions that include precise information about their singularities. We instead independently estimate $f_{ijkl}^{\textrm{aniso}}$ by examining the largest system sizes of simulations performed at $r=1.2, 1.5$. In addition, we analyze the data for $Y=0.66, 1.65$. 

We find that the critical exponents for $C_{xxxx}$ are distinct from those found for the isotropic system while those for the other elasticity tensor components cannot be distinguished from those found for the isotropic system:
\beq
    \begin{aligned}
        f^{\textrm{aniso}}_{xxxx} &\neq f^{\textrm{iso}}_{xxxx} \\     f^{\textrm{aniso}}_{ijkl} &\approx f^{\textrm{iso}}_{ijkl} & (ijkl \neq xxxx) \\
    \end{aligned}
\eeq
with $f^{\textrm{aniso}}_{xxxx} = \faniso \pm \fanisoerr$, compared to our estimate of $f^{\textrm{iso}} = \fiso \pm \fisoerr$. 
As usual in crossover scaling, the multicritical point governs short
length scales and regions far from the critical lines emanating from it.
We thus expect to find a crossover from $f^{\textrm{aniso}}_{xxxx}$ to $f^{\textrm{iso}}_{xxxx}$ as networks move away from the anisotropic critical point, as demonstrated in Fig.~\ref{fig:CrossoverP}. We also attempt to independently estimate $\nu^{\textrm{aniso}}$, the finite-size scaling exponent away from isotropy, using the collapse plots shown in Fig.~\ref{fig:SigmaCollapse} (Appendix~\ref{AppAnisoCollapse}), but our estimates span a wide range of values $1.2-3.2$. Our estimates for all scaling exponents are shown in Table~\ref{table:Exponents}.

\begin{table}[H]
\centering
    \begin{ruledtabular}
    \begin{tabular}{lr} \textbf{Exponent} & \textbf{Estimate} \\
        \hline
         $\nu$                & $\nue \pm \nuerr$ \\
         $f^\textrm{iso}$     & $\fiso \pm \fisoerr$ \\
         $\zeta$              & $\zetae \pm \zetaerr$ \\
         $\nu^{\textrm{aniso}}$ & $1.2-3.2$ \\
         $f_{xxxx}^\textrm{aniso}$   & $\faniso \pm \fanisoerr$ \\
         $f_{ijkl}^\textrm{aniso}$ ($ijkl \neq xxxx$)   & $\fiso \pm \fanisoerr$ 
         \end{tabular}
    \end{ruledtabular}
    \caption{\textbf{Numerical estimates of critical exponents.}}\label{table:Exponents}
\end{table}

\section{Discussion}

We find that rigidity percolation in our model anisotropic system occurs in at least two steps, with the modulus in the direction of alignment becoming nonzero at lower volume fractions. Our estimate of at least one of the critical exponents of the anisotropic transition, $f^{\textrm{aniso}}_{xxxx}$, appears distinct from the corresponding exponent for the isotropic transition $f^\textrm{iso}$, which suggests that these anisotropically diluted networks feature two distinct universality classes.

There are several differences between the predictions of a simple effective medium theory (see~\cite{AnisotropicSpringNetworks} and our reproduction in Appendix~\ref{EMTApp}) and our detailed numerical analysis of this model. The effective medium theory predicts a phase boundary that is consistent with a simple Maxwell counting estimate, which would correspond to a vertical line of transitions for all moduli at $p=2/3$ in Fig.~\ref{fig:PhaseDiagram}. It also predicts a mean-field-like $f=1$ critical exponent both at and away from isotropy. In contrast, our numerical simulations of the model suggest violation of the simple application of Maxwell counting, an entirely new \textit{phase} that is seen in the thermodynamic limit, and nontrivial exponents both at and away from isotropy.

It was a surprise to us that the rigidity percolation transition for the isotropic lattice broke up into several transitions when it became anisotropic. First, obtaining multiple transitions is contrary to the na\"ive usage of Maxwell counting to determine the location of the rigidity transition. As shown in the insets of Fig.~\ref{fig:PhaseDiagram}, the rigid modes become anisotropic, and span horizontally before they span vertically.

Second, our results are fundamentally different from connectivity percolation, where regardless of the value of $r$ there can only be one transition. 
The left inset in Fig.~\ref{fig:PhaseDiagram} shows several horizontal stress-supporting chains spanning the network. The critical point at which $C_{xxxx}$ first becomes non-zero presumably separates a phase where there are no stress-supporting chains from one where there are a finite density of such chains. In regular percolation, two such paths connecting the system horizontally that are separated by any finite distance will have a finite probability per unit length of being connected by bonds extending in the vertical direction. Hence, for ordinary percolation, as soon as one crosses the horizontal percolation point in an anisotropic system, it must percolate in the other directions as well. This argument suggests that lattices with bending stiffnesses and angular springs, which are believed to become rigid at the connectivity percolation threshold~\cite{RedundancyCrossLink}, lack this intermediate phase. However, in typical situations where bending stiffnesses are much weaker than stretching stiffnesses, a remnant of this intermediate phase should be measurable even when bending is included.

In retrospect, we should have expected separate transitions in central force rigidity percolation. Maxwell counting tells us when the number of zero modes can vanish in the absence of states of self stress, but does not tell us whether the zero modes couple to a given mode of deformation. Clusters supporting horizontal stress in a large system, when connected vertically, may only contain contributions to the stress that grow quadratically (i.e. non-linearly) in the strain: the length of a beam connecting $(x, y)$ to $(x', y+\epsilon)$ grows as $\epsilon^2$, suggesting the corresponding linear elastic modulus is $0$. A similar nonlinear response to infinitesimal strains is found to stabilize hypostatic jammed packings of ellipsoidal particles~\cite{HypostaticPacking2007}, in violation of simple constraint-counting arguments. The perfect square lattice has no $C_{xyxy}$ shear modulus, and the perfect hexagonal lattice has no non-zero moduli except the bulk modulus -- why should anisotropic random lattices not possess separate transitions? Along these lines it might be interesting to investigate whether separate rigidity transitions arise in floppy isotropic random lattices under finite deformation. Here, strain stiffening could lead to rigidity along the extension axis. For example, stretching a non-disordered hexagonal lattice until two of the three bond orientations become parallel lines of bonds would yield a non-disordered form of our $C_{xxxx} > 0$ phase, with every horizontal bond connected at its two ends to the next row vertically.

What kind of rigidity critical points do we expect? The $C_{xxxx}$ transition where horizontal stress-bearing chains first arise could be self-similar (with a single diverging correlation length), but could also be self-affine (with the vertical spacing between chains diverging with a different power on the stress-supporting side than the rigid cluster lengths diverge on the floppy side). Whether the three other moduli become non-zero simultaneously or separately in this model is not numerically resolved yet, but one expects that more complicated anisotropies (say a 3D model with brick-like symmetry) will allow for separate transitions for those moduli as well.

Finally, it would be interesting to consider whether the results presented here have any bearing on biological networks. In particular, there are many situations where intercellular cytoskeletal networks and extracellular networks show preferential alignment, either due to morphogenesis or due to external loads straining the networks. While a quantitative comparison may require analyzing the effects of second-order constraints, changing the contact number distribution, and exploring the crossover between this pair of zero-bending stiffness transitions and a bending-dominated regime, the qualitative effect of the splitting of the phase transition may be able to be observed in real systems. Moreover, cells are able to control how matrix elements are generated. Cells may generate networks with many different rigidity transitions to tune between, where the particular way matrix elements are laid down biases the thresholds of the different components of the elasticity tensor. As such, the results presented here could have profound implications for understanding more complicated networks in many biological systems.

\acknowledgments

The authors acknowledge Thomas Wyse Jackson for critical discussions in the early phases of the project. WYW, IC, NS, SJT, and JPS are supported by NSF DMR 2327094. JO was supported in part by the NSF MRSEC DMR-1719875. BC was supported by NSF CBET award number 2228681 and NSF DMR award number 2026834. AB is supported by NSF DGE 2139899. JAM and MD are supported by NSF EF 1935277.

\bibliographystyle{apsrev4-1_custom}
\bibliography{refs}

\appendix
\counterwithin{figure}{section}  %
\renewcommand*\thefigure{\thesection\arabic{figure}} %

\section{Effective Medium Theory}
\label{EMTApp}

We analyze the results of a previously-utilized effective medium theory for bond-diluted lattices with central-force springs, reproducing the calculation of Zhang et al.~\cite{AnisotropicSpringNetworks}. We compare this with our numerical findings for both the shape of the phase boundary found in Fig.~\ref{fig:PhaseDiagram}(b) and the power-law vanishing of the modulus with $\left|\delta p\right|^{f^{\textrm{iso}}}$ found through a finite-size scaling analysis shown in Fig.~\ref{fig:UniversalIso}.

A triangular lattice with net bond occupation probability $p$ and anisotropy parameter $r\geq1$ has bonds along its three directions $\left\{x,y,z\right\}$ with probabilities $\left\{3pr/\left(2+r\right),\;3p/\left(2+r\right),\;3p/\left(2+r\right)\right\}$.\footnote{These probabilities can be found by solving ${\left(p_x+p_y+p_z\right)/3=p},\;{p_x/p_y=r},\; {p_y=p_z}$ for $\left\{p_x,p_y,p_z\right\}$ in terms of $p$ and $r$. Not all values of $0\leq p\leq1$ and $0\leq r\leq \infty$ are possible; the other boundaries $p_x=1$ and $p_y=1$ get warped to the curves $r_{\textrm{max}}(p)=2/(3p-1)$ and $r_{\textrm{min}}(p)=3p-2$, respectively. For large $p$, the lattice cannot be too anisotropic.} We assume that each bond is independently populated according to the bimodal distribution
\begin{equation}
P\left(k_{\alpha}'\right) = p_{\alpha}\,\delta\left(k_{\alpha}'-1\right)+\left(1-p_{\alpha}\right)\,\delta\left(k_{\alpha}'\right)
\end{equation} where $\alpha\in\left\{x,y,z\right\}$. The effective medium theory assumes that the randomly diluted triangular lattice at parameters $(p,r)$ can be well-represented by a completely filled triangular lattice with renormalized stiffnesses in each direction. To self-consistently calculate the renormalized stiffnesses, one writes the Green's function for a phonon in the effective medium ($\mathbf{G}$), and perturbs it by a random bond in one of the sublattices (giving $\mathbf{G}^V$). This perturbation amounts to
\begin{equation}
\mathbf{G}^V=\mathbf{G}+\mathbf{G}\mathbf{V}\mathbf{G}+\mathbf{G}\mathbf{V}\mathbf{G}\mathbf{V}\mathbf{G}+\dots = \mathbf{G} + \mathbf{G}\mathbf{T}\mathbf{G}
\end{equation}
where $\mathbf{V}$ is the scattering potential introduced by the random bond, and $\mathbf{T}$ is the scattering $T$-matrix. One then requires that the average scattering from the perturbed lattice is equivalent to the scattering from the effective medium. Since the random perturbing bond only enters through $\mathbf{T}$, this amounts to the condition that
\begin{equation}
\left\langle\mathbf{T}\right\rangle = p_{\alpha}\mathbf{T}^{\alpha}_{k_{\alpha}=1}+\left(1-p_{\alpha}\right)\mathbf{T}^{\alpha}_{k_{\alpha}=0}=0.
\end{equation}

In our system, there is no distinction between $y$ and $z$ bonds, so we can treat this as a two-sublattice system with two coupled equations that are simultaneously solved for the effective $k_{x}$ and $k_{y/z}$ after the lattice has been partially diluted. The full dynamical matrix for a triangular lattice is decomposed into one for $x$ bonds $\mathbf{D}_{x}=k_{x}\mathbf{K}_{x}$ and one for $y/z$ bonds $\mathbf{D}_{y/z}=k_{y/z}\mathbf{K}_{y/z}$. The self-consistent equations for the stiffnesses $k_{\alpha}$ are
\begin{multline}
\widetilde{z}_{\alpha}\frac{p_{\alpha}-k_{\alpha}}{1-k_{\alpha}}=\\
=\frac{1}{s_{\textsc{BZ}}}\int_{1\textrm{BZ}}\textrm{d}^2\mathbf{q}\;\textrm{Tr}\left(k_{\alpha}\mathbf{K}_{\alpha}\left(\sum_{\beta}k_{\beta}\mathbf{K}_{\beta}\right)^{-1}\right)
\label{eqn:sceq}
\end{multline}
where $\alpha=x$ or $y/z$, $\widetilde{z}_{\alpha}$ is the number of bonds per site in the $\alpha$ sublattice, and $s_{\textsc{BZ}}$ is the volume of the first BZ. On the floppy side of the transition, the full dynamical matrix fails to be invertible. This can be cured by extending the theory to finite frequency and then taking the zero frequency limit, equivalent to the regularization
\begin{multline}
\widetilde{z}_{\alpha}\frac{p_{\alpha}-k_{\alpha}}{1-k_{\alpha}}=\\
=\frac{1}{s_{\textsc{BZ}}}\int_{1\textrm{BZ}}\textrm{d}^2\mathbf{q}\;\textrm{Tr}\left(k_{\alpha}\mathbf{K}_{\alpha}\left(\sum_{\beta}k_{\beta}\mathbf{K}_{\beta}-\varepsilon \mathbf{I}\right)^{-1}\right) 
\end{multline}
with $\varepsilon\rightarrow 0^+$. 

For the triangular lattice with lattice vectors $\mathbf{a}_1=\left(1,0\right)$ and $\mathbf{a}_2=\left(1/2,\sqrt{3}/2\right)$ we have the following stiffness matrices:
\begin{equation}
\mathbf{K}_{x}=\begin{pmatrix}
4\sin^2\left(q_x/2\right) & 0 \\
0 & 0
\end{pmatrix}
\end{equation}
and
\begin{multline}
\mathbf{K}_{y/z}\\
=\begin{pmatrix}
1-\cos\left(\frac{q_x}{2}\right)\cos\left(\frac{\sqrt{3}q_y}{2}\right)& \sqrt{3}\sin\left(\frac{q_x}{2}\right)\sin\left(\frac{\sqrt{3}q_y}{2}\right) \\
\sqrt{3}\sin\left(\frac{q_x}{2}\right)\sin\left(\frac{\sqrt{3}q_y}{2}\right) & 3-3\cos\left(\frac{q_x}{2}\right)\cos\left(\frac{\sqrt{3}q_y}{2}\right)
\end{pmatrix}.
\end{multline}
The first Brillouin zone is a hexagon with volume ${s_{\textsc{BZ}}=8\pi^2/\sqrt{3}}$.

First, it can be noted that summing the two self-consistent equations for $x$ and $y/z$ (Equation~\ref{eqn:sceq}) leads to the sum rule (on the solid side)
\begin{equation}
\widetilde{z}_{x}\frac{p_{x}-k_{x}}{1-k_{x}}+\widetilde{z}_{y/z}\frac{p_{y/z}-k_{y/z}}{1-k_{y/z}}=2
\end{equation}
as the integrand becomes the trace of the $2\times2$ identity matrix. The locations where the stiffnesses first vanish can be found by simply setting $k_{x/y}=k_z=0$, since this is equivalent to approaching the phase boundary from the solid side. This gives a phase boundary
\begin{equation}
\widetilde{z}_{x}p_{x}+\widetilde{z}_{y/z}p_{y/z}=p_{x}+2p_{y/z}=2.
\end{equation}
In the language of our model with $p$ and $r$, this translates to
\begin{equation}
\frac{3p_cr}{2+r}+2\frac{3p_c}{2+r}=2\implies p_c=\frac{2}{3},
\end{equation}
independent of $r$. This is a statement of the Maxwell counting constraint, that the average number of constraints $\widetilde{z}p=3p$ is equal to the number of degrees of freedom $d=2$, which is unaffected by $r$. Even in the isotropic case, the reported value of $p_c$ from numerical studies is typically found \cite{CriticalityIsostaticity} to be smaller than $2/3$, consistent with our numerical results. We also find from our simulations that our phase boundaries for the different independent moduli bend in different ways as we move to $r>1$, leading to a \textit{pair} of rigidity transitions and \textit{three} distinct phases. This behavior is completely uncaptured by the effective medium theory.

We can also analyze the critical exponents predicted by this theory. By expanding Equation~\ref{eqn:sceq} close to the critical point in powers of $\delta p$, we find that both $k_x$ and $k_{y/z}$ vanish linearly with $\delta p$ upon approaching the phase boundary ($k\sim\left|\delta p\right|^1$) away from the point where the phase boundary intersects the line $r=2$. In the long-wavelength limit, all of the independent components of the elasticity tensor are proportional to linear combinations of $k_{x}$ and $k_{y/z}$, and so the effective medium theory predicts that all $C_{ijkl}\sim\left|\delta p\right|^1$, clearly at odds with our numerical findings at and away from isotropy.

The point in the phase diagram of the effective medium theory $p=2/3$, $r=2$ plays the role of a multicritical point of the same character as the one studied in \cite{LiarteLub2019} (and is equivalent to the point $p_x=1$, $p_y=1/2$ in \cite{AnisotropicSpringNetworks}), where $k_{x}$ jumps discontinuously to a finite value as we approach from the floppy phase due to the formation of system-spanning chains of horizontal bonds. Our analysis of our numerical simulation is performed as a crossover scaling between transitions in the vicinity of the isotropic rigidity percolation point $r=1$, so we expect the effects of the other distinct predicted transition at $r=2$ (far away in the phase diagram) to enter as analytic corrections to scaling.

\section{Bond filling protocol}
\label{BondFilling}

One method for filling the lattice is to choose some number of bonds $n$ to randomly occupy, and set $p=n/N$, where $N$ is the total number of possible bonds. This has the disadvantage that changes in $p$ can only be measured to a sensitivity $1/N$. To characterize the behavior very close to the critical point, we instead fill our lattice in a way that is statistically equivalent, but allows measurements at continuous values of $p$. In the isotropic case, the algorithm is as follows: first, a random number $s_i$ taken uniformly between 0 and 1 is assigned to each bond $i$. At a filling parameter value $p$, all bonds $i$ with assigned random numbers $s_i<p$ are filled, and the independent components of the linear elasticity tensor are measured through applied shears. For different random number seeds, the ``jumps’’ in the linear moduli associated with the addition of single stress-supporting bonds to the rigid backbone occur at different values of $p$ (which are not multiples of $1/N$). When the measurements are averaged over several random number seeds, we find that the measurements of moduli quickly converge to a smooth function of $p$ at a given system size $L$, except at the smallest values of $p$. This algorithm is modified to include our anisotropy parameter $r$ in a straightforward way.

We start by picking a random number seed, and then assigning a (uniformly chosen) random number $s_i$ between 0 and 1 to each bond. The bonds are assigned and then sorted based on a key, $k_i$, which corresponds to the value of bond occupation fraction $p$ for which the bond would be added based on the anisotropy parameter $r$:
\begin{equation}
    k_i^x = \frac{(2+r)s_i^x}{3r},\quad  k_i^y = \frac{(2+r)s_i^y}{3},
    \label{eqn:BondKeys}
\end{equation}
where $s_i^x$ are assigned to bonds in the horizontal direction and $s_i^y$ are assigned to the other bonds. The bonds are then removed according to their keys, highest to lowest.
Note that this formulation allows for finding the bond configuration while continuously varying both $p$ and $r$.

\section{Numerical Methods}\label{AppNumerics}

We introduce strains to the lattice by applying the proper transformation matrix to the positions of each of the nodes. For example, to stretch the network in the horizontal direction (i.e., to apply strain $\varepsilon_{xx}$), the following matrix is applied:
\beq
    T = \begin{bmatrix}
        1 + \gamma & 0 \\ 
        0 & 1
    \end{bmatrix},
\eeq
where we set $\gamma$ to $10^{-3}$. Since we implement periodic boundary conditions in both directions, all nodes are transformed in the same manner. 

We then minimize the energy given in Equation~(\ref{eqn:elasticTensor}), which can be equivalently written as  
\beq
    E = \frac{1}{2} \vec{u}^\top H \vec{u}, \quad H_{ij} = \frac{\partial^2 E}{\partial u_i \partial u_j},
\eeq
where $H$ is the Hessian matrix and $\vec{u}$ is a length $N \times d$ vector containing the displacements from the initial node position.

To handle periodic boundary conditions, we split the Hessian into two parts: $H_{\text{pbc}}$, which is computed using only the bonds that span across the network, and $H_{\text{in}}$ with bonds which do not. The energy is therefore computed as:
\beq
    E = \frac{1}{2} \vec{u}^\top H_{\text{in}} \vec{u} + \frac{1}{2} (\vec{u} + \vec{c})^\top H_{\text{pbc}} (\vec{u} + \vec{c}),
\eeq
in which $\vec{c}$ ``corrects'' the displacements for nodes that are connected across the network and depends on the particular strain. The energy is then minimized by finding a zero-force configuration, solving the following linear system:
\beq
    (H_{\text{in}} + H_{\text{pbc}})\vec{u}_{\text{relaxed}} = -H_{\text{pbc}}\vec{c}
\eeq

An affine displacement is used as an initial guess. The sparsity structure allows matrices to be stored in compressed sparse row format, reducing memory usage and improving the speed of operations. To improve convergence rates, an incomplete Cholesky factorization for $(H_{\text{in}} + H_{\text{pbc}})$ is computed and used as a preconditioner for a conjugate gradient method \cite{ConjugateGradientMethods}. For large system sizes, we use GPUs to accelerate numerical computations, such as matrix factorizations and matrix-vector products.

We note that for systems that have under-constrained nodes, the null space of $(H_{\text{in}} + H_{\text{pbc}})$ has a non-zero dimension; as such we do not consider the non-affinity parameter (which sums the squared displacements from an affine transformation) as a method of analysis or for extracting critical exponents.

\section{Details of finite-size effects}\label{AppDistCollapse}

To obtain a consistent estimate of $p_c^\infty$ and $\nu$, we consider the distribution of the rigidity percolation threshold for each modulus at isotropy ($r=1$). We take $p_c^{ijkl}$ to be the smallest value of $p$ at which $C_{ijkl}$ is rigid. For a given system size $L$, the value of $p_c^{ijkl}$ is sampled from an underlying distribution:
\beq
    p_c^{ijkl}(L) \sim \rho^{ijkl}_L, \quad \rho^{ijkl}_L \in \Delta[0,1].
\eeq

\begin{figure}[!ht]
\begin{center}
\includegraphics[alt={isotropic percolation threshold histograms}, width=\linewidth]{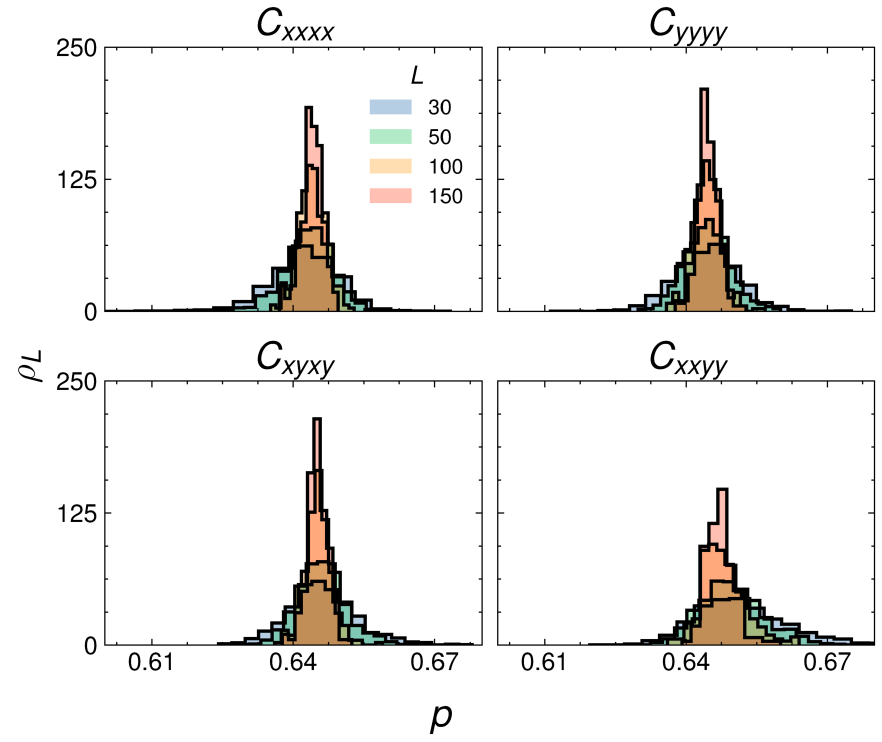}
\end{center}
\caption{\textbf{Histograms of rigidity percolation threshold $p_c$ as a function of system size $L$ at isotropy.} The estimated density functions for each independent elastic modulus are plotted. Each distribution becomes increasingly sharp with larger system size.}
\label{fig:CDistribution}
\end{figure}

Histograms of $p_c^{ijkl}$ are plotted in Fig.~\ref{fig:CDistribution}. As the system size increases, the distributions become increasingly sharp and the means shift systematically. In the limit of infinite system size, we assume that each distribution converges to a delta function about $p_c^\infty$.

For each system size, we compute both the means $\langle p_c^{ijkl} \rangle_L$ and standard deviations $\sigma^{ijkl}_L$ of each distribution. We expect systematic shifts in the means and standard deviations (i.e., the first and second moments) to scale as a power law with respect to $L$ governed by a single critical exponent $\nu$:
\beq
\begin{aligned}
    \label{eqn:finite-size}
    \langle p_c^{ijkl} \rangle_L - p_c^\infty &\sim L^{-1/\nu} \\ 
    \sigma^{ijkl}_L &\sim L^{-1/\nu}
\end{aligned}
\eeq

\begin{figure}[!ht]
\begin{center}
\includegraphics[alt={mean and standard deviation shifts}, width=\linewidth]{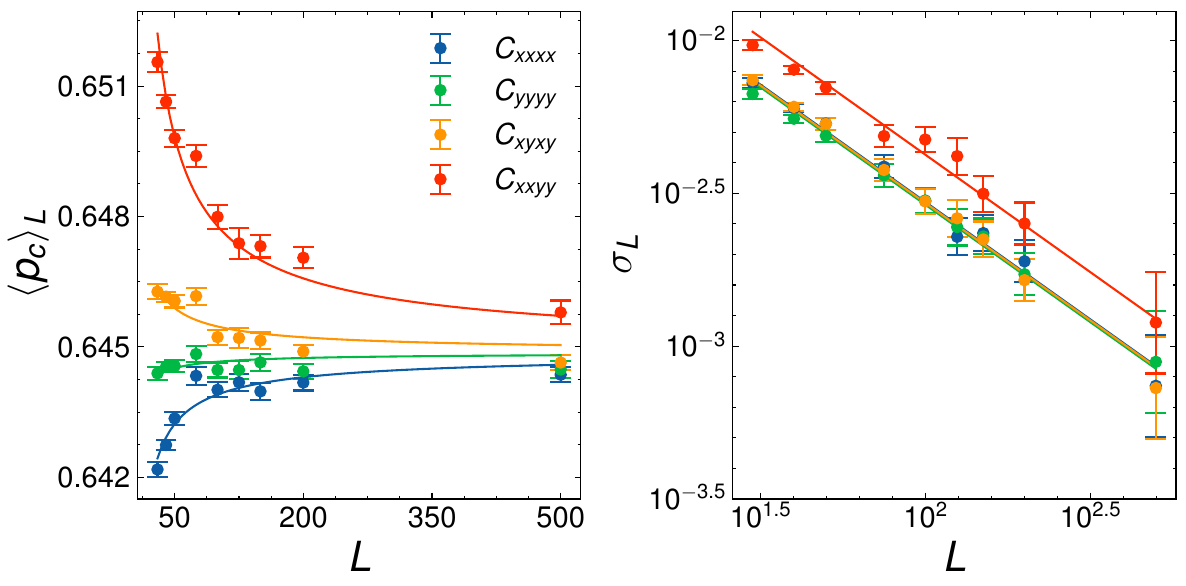}
\end{center}
\caption{\textbf{(left) Mean of $p_c$ and (right) standard deviations of the $p_c$ distribution at isotropy as a function of system size.} The fitted curves match those in Equation~\ref{eqn:finite-size} and the fitted lines on the right figure have slope $-1/\nu$.}
\label{fig:C_sigma}
\end{figure}

We perform a joint non-linear least squares fit  \cite{LMFIT}, with $p_c^\infty$ and $\nu$ the same for all curves (we assume they are equal for each modulus at isotropy). Figure~\ref{fig:C_sigma} depicts $\langle p_c^{ijkl} \rangle_L$ (left) and $\sigma^{ijkl}_L$ (right) as a function of $L$. The fits give an estimate of $p_c^\infty = \pcinfty \pm \pcinftyerr$ and $\nu = \nue \pm \nuerr$.

\begin{figure}[!ht]
\begin{center}
\includegraphics[alt={percolation histograms collapse}, width=\linewidth]{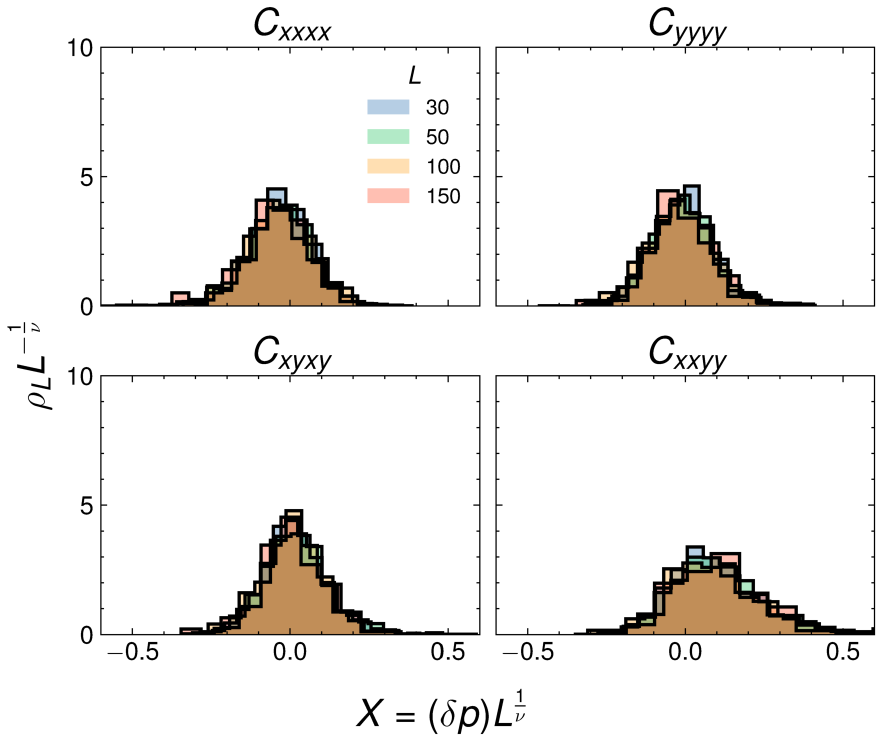}
\end{center}
\caption{\textbf{Universal scaling of rigidity distributions at isotropy for each independent modulus.} The histograms all collapse when plotted against the scaling variable $X$.}
\label{fig:CCollapse}
\end{figure}

Furthermore, we find a universal scaling function for the distributions $\rho^{ijkl}_L$ with respect to our previously defined scaling variable $X \equiv (\delta p) L^{1/\nu}$:
\beq
    \label{eqn:dist_collapse}
    \rho^{ijkl}_L(p) \sim L^{1/\nu} \mathcal{R}_{ijkl} (X).
\eeq
We find that $X$ collapses the density functions, with the resulting histograms shown in Fig. \ref{fig:CCollapse}.

\section{Details of modulus scaling collapse}\label{AppIsoModulusCollapse}

\begin{figure}[!ht]
\begin{center}
\includegraphics[alt={unscaled modulus data}, width=\linewidth]{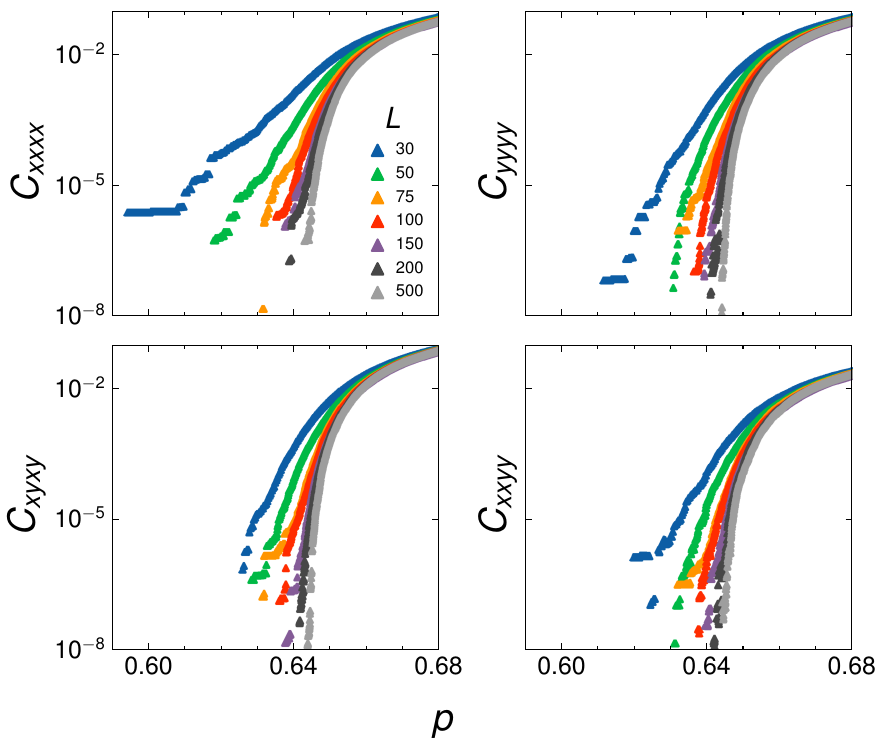}
\end{center}
\caption{\textbf{Unscaled modulus data across various system sizes at isotropy $(r=1)$.} The smaller system sizes have a higher probability of becoming rigid at lower values of $p$.}
\label{fig:AllModuliUncollapse}
\end{figure}

At isotropy, each modulus grows as a power law above rigidity percolation threshold $C_{ijkl} \sim (\delta p)^{f^\textrm{iso}}$. Figure~\ref{fig:AllModuliUncollapse} depicts the uncollapsed finite-size scaling data at isotropy; the systematic deviations are clear.

\begin{figure}[!ht]
\begin{center}
\includegraphics[alt={isotropic scaling function of all moduli}, width=\linewidth]{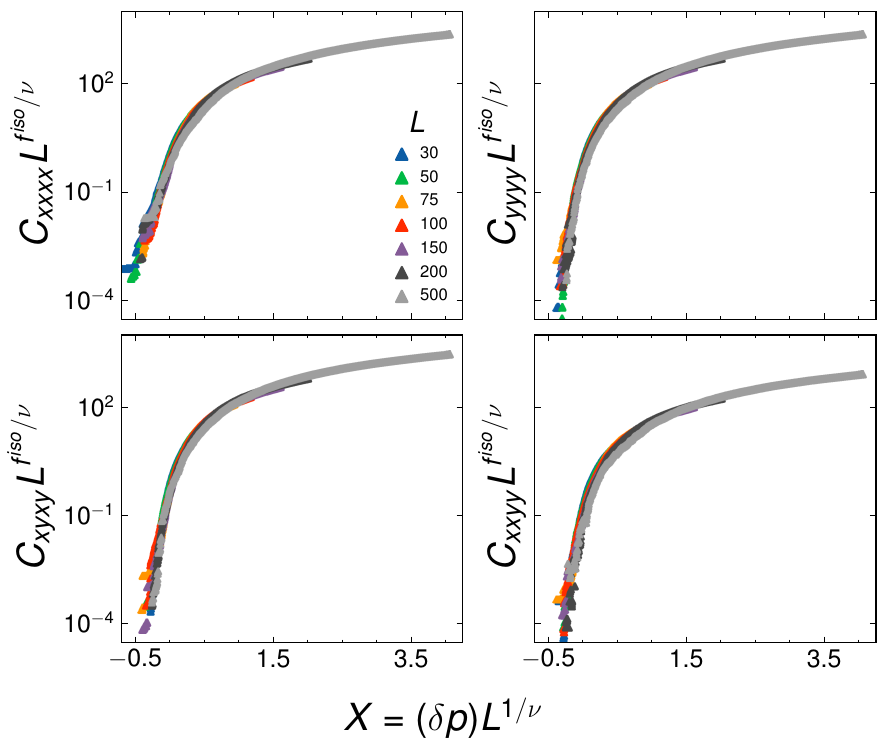}
\end{center}
\caption{\textbf{Universal scaling function of all independent elastic moduli at isotropy ($r=1$).} The independent components of the elasticity tensor each collapse onto a single curve $\mathcal{C}_{ijkl}^{\textrm{iso}}$ when plotted against the finite-size scaling variable $X \equiv (\delta p)  L^{1 / \nu}$. In this isotropic case, there are only two independent moduli in the long-wavelength elasticity tensor ($B$ and $G$, for instance).}
\label{fig:AllModuliCollapse}
\end{figure}

We find the exponent $f^{\textrm{iso}}$ by considering the largest available system size ($L=500$) and performing a least-squares fit of the modulus, finding a range of $f^{\textrm{iso}}$ as we vary $p_c$ slightly. With $f^{\textrm{iso}} = \fiso$, $p_c^\infty = \pcinftycollapse$, and $\nu = \nue$, we plot the data against the proposed scaling variables and find a nice collapse for all independent moduli, shown in Fig.~\ref{fig:AllModuliCollapse}. The number of samples we average over ranges from $10^4 \mbox{--} 10^2$ for system sizes $L \in [30, 200]$ and $20$ samples of $L=500$.

\begin{figure}[!ht]
\begin{center}
\includegraphics[alt={isotropic scaling function of all moduli with previous exponents}, width=\linewidth]{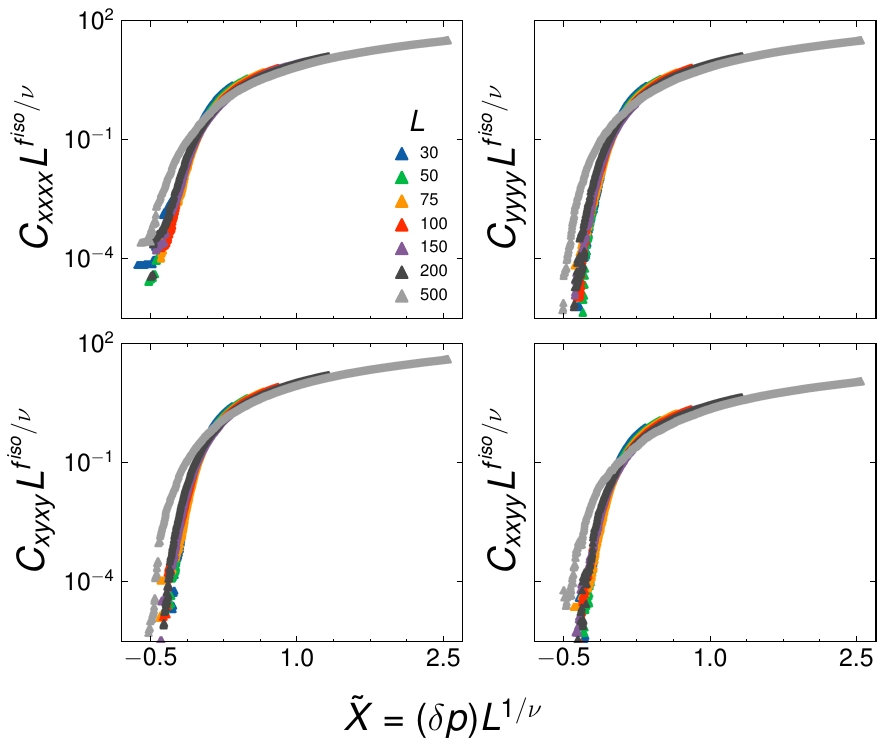}
\end{center}
\caption{\textbf{Universal scaling function of all independent elastic modulus using previously reported exponents~\cite{CriticalityIsostaticity}}. The data for the range $L \in [30, 200]$ comparable to the earlier work does give a good collapse. Having the larger system size ($L=500$) explains why we find different exponents.}
\label{fig:ModuliCollapseMackintosh}
\end{figure}

Using the value of $f^{\textrm{iso}} = 1.4 \pm 0.1$ and $\nu = 1.4 \pm 0.2$ quoted in \cite{CriticalityIsostaticity}, we find best collapse with $p_c^\infty = 0.65$ shown in Fig.~\ref{fig:ModuliCollapseMackintosh}, which gives good collapse for lower values of system size, but does not collapse the modulus for our largest system. 

\section{Separation of phase transitions}\label{AppPhaseSeparation}

\begin{figure}[!ht]
\begin{center}
\includegraphics[alt={anisotropic percolation threshold histograms}, width=\linewidth]{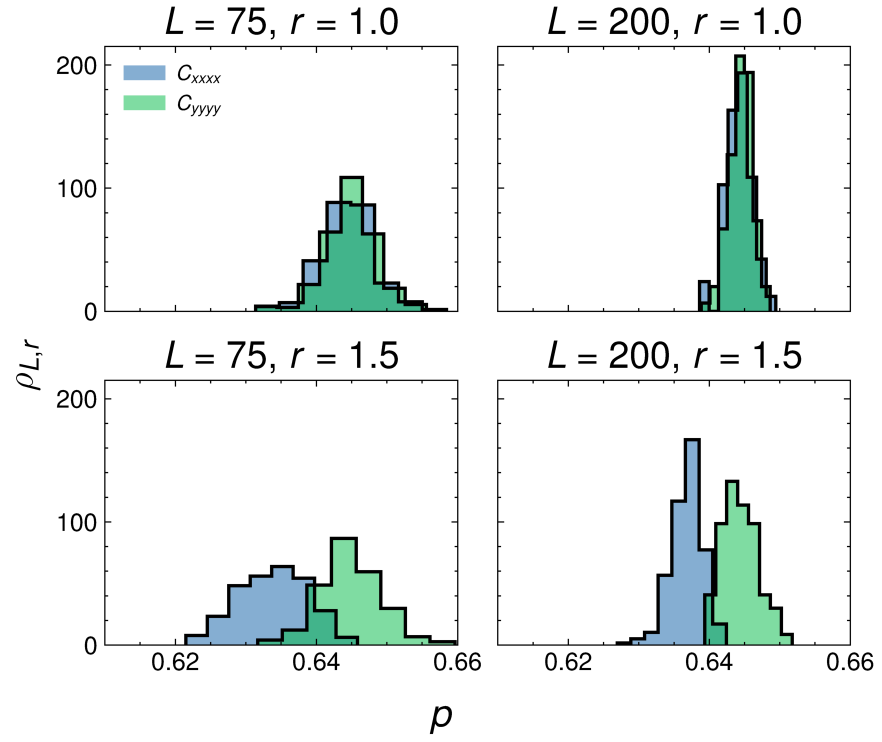}
\end{center}
\caption{\textbf{Histograms of rigidity percolation threshold $p_c$ for $C_{xxxx}$ (blue) and $C_{yyyy}$ (green) for the same system size.} The means are consistent at isotropy ($r=1$) and distinct from each other at higher anisotropy ($r=1.5$).}
\label{fig:SeparationDistribution}
\end{figure}

Here we present our numerical evidence for the separation of the two phase transitions (one for $C_{xxxx}$, and at least one additional for $C_{ijkl}$ with $ijkl\neq xxxx$) as $L\rightarrow\infty$. We do this by analyzing the systematic dependence on $L$ of the distributions of $p_c$ for each modulus away from isotropy, i.e., we perform large numbers of simulations at various system sizes for fixed $r=1.5$ (away from isotropy) and examine how these distributions depend upon $L$. An example of these distributions can be seen in Fig.~\ref{fig:SeparationDistribution}, where at $L=75$ (first column) the distributions of $p_c$ for the $C_{xxxx}$ and the $C_{yyyy}$ moduli have significant overlap, but when we look at $L=200$ (second column) the distributions are beginning to separate for $r>1$.

Sample-to-sample, there are lattices that can support rigidity in some shear directions but not others. There are two basic scenarios. (1) In the case where including anisotropy ends up simply giving analytic corrections to scaling that bend a single phase boundary, all moduli will vanish at the same location at $L=\infty$, but the amplitudes of the finite-size effects may be different. Singling out the $C_{yyyy}$ modulus for the sake of comparison, this means that
\beq
\left\langle p_{c}^{ijkl}\right\rangle_L -\left\langle p_{c}^{yyyy}\right\rangle_L \sim L^{-1/\nu}.
\eeq
This is the same as the asymptotic scaling for the spreads of these distributions, $\sigma^{ijkl}_L \sim L^{-1/\nu}$, as all distributions are controlled by the exponents of the isotropic critical point in this supposition. If we measure the separation between the means as a function of system size in terms of the number of standard deviations of the distribution at that system size, using the more democratic ${\sigma_L^2=(\sigma^{ijkl}_L)^2+(\sigma^{yyyy}_L)^2}$, then, we should find
\beq
\frac{\left\langle p_{c}^{ijkl}\right\rangle_L -\left\langle p_{c}^{yyyy}\right\rangle_L }{\sqrt{ ({\sigma^{ijkl}_L})^2+ (\sigma^{yyyy}_L)^2}}\sim c_{ijkl},
\eeq
for some constant $c_{ijkl}$, which is flat as a function of system size.

(2) In the case where including anisotropy leads to genuinely new critical phenomena and a pair of phase transitions, the finite-size effects of the mean and standard deviation are controlled by the finite-size scaling exponent of each anisotropic rigidity transition $\nu^{\textrm{aniso}}$. If we split into a pair of phase transitions, then the spreads $\sigma_{ijkl}$ of each distribution will narrow with increasing $L$, but the separation of the means is asymptotically constant as $L\rightarrow\infty$. This would make
\beq
\frac{\left\langle p_{c}^{ijkl}\right\rangle_L -\left\langle p_{c}^{yyyy}\right\rangle_L }{\sqrt{ ({\sigma^{ijkl}_L})^2+ (\sigma^{yyyy}_L)^2}}\sim L^{1/\nu^{\textrm{aniso}}},
\eeq
where $\nu^{\textrm{aniso}}$ is the largest of the (potentially different) finite-size scaling exponents associated with the new, anisotropic transition.

\begin{figure}[!ht]
\begin{center}
\includegraphics[alt={separation of percolation threshold means}, width=\linewidth]{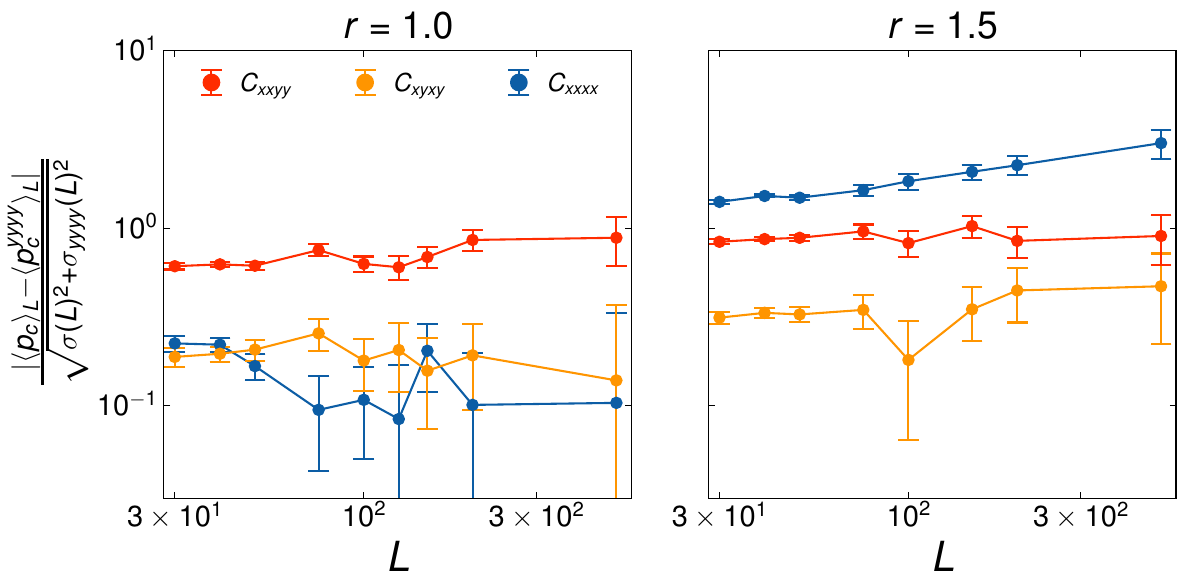}
\end{center}
\caption{\textbf{Separation of rigidity percolation threshold mean $\langle p_c \rangle$ from that of the $C_{yyyy}$ modulus as a function of system size at isotropy (left) and anisotropy (right).}}
\label{fig:SplittingPhase}
\end{figure}

We begin by performing this measurement of the separations between the distributions of $p_c$ for all moduli at the isotropic transition, where all moduli vanish at the same location in $p$ as $L\rightarrow\infty$ (Fig.~\ref{fig:SplittingPhase} (left)).

As expected, this measure of the separation in the means is flat as a function of system size, confirming that these moduli vanish at the same asymptotic location and that the finite-size effects controlling the mean and the standard deviation have the same systematic dependence on $L$.

When we perform the same analysis for the anisotropic case $(r=1.5)$, we see systematic growth in this measure as a function of system size (Fig.~\ref{fig:SplittingPhase} (right)), suggesting that the locations of $p_c$ for different moduli are genuinely different in the thermodynamic limit. This is moderately strong quantitative evidence for the information that can roughly be seen by eye in Fig.~\ref{fig:SeparationDistribution}.

\section{Estimate of anisotropic scaling exponents}\label{AppAnisoCollapse}

At infinite system size, the phase diagram curves in Fig.~\ref{fig:PhaseDiagram}
contain important information about the critical exponent $\zeta$ near the isotropic transition.
We find $\zeta$ by fitting the differences between the $C_{xxxx}$ curves and the $C_{ijkl}$ curves (where $ijkl \neq xxxx$) to a power law, resulting in a value of $\zeta = \zetae \pm \zetaerr$. The individual phase boundaries have an important linear correction to scaling, as the unstable eigenvector is not along the $r$-axis, but has a slope $m$. Hence the two phase boundaries are of the form $\delta p = m (r-r_c) + W (r-r_c)^{1/\zeta}$, with a fixed value of $W$ defining a curve along which the invariant scaling combination is constant. Because $\zeta$ is small, this correction cannot be neglected. By fitting the differences between the phase boundaries, we bypass this linear correction to scaling.

\begin{figure}[!ht]
\begin{center}
\includegraphics[alt={scaling collapse of rigidity distributions}, width=\linewidth]{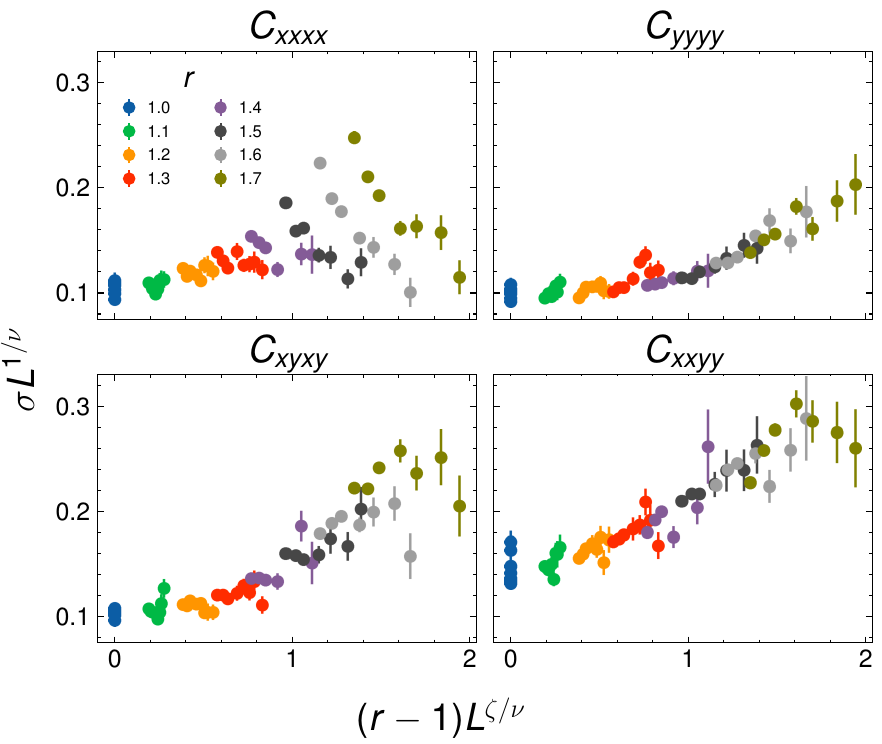}
\end{center}
\caption{\textbf{Universal scaling of rigidity distributions near isotropy for each independent modulus.} The widths of the histograms all collapse onto a single curve when plotted against the finite-size scaling variable $Y$. There appear to be deviations in the collapse of the $C_{xxxx}$ modulus at higher values of $Y$.}
\label{fig:SigmaCollapse}
\end{figure}

Furthermore, we can consider the standard deviations of the rigidity percolation threshold distributions (as in Appendix~\ref{AppDistCollapse}) for $r>1$ and consider a scaling function with respect to our scaling variable $Y = (r-1)L^{\zeta / \nu}$
\beq
    \sigma^{ijkl}(L, r) \sim L^{-1/\nu} \mathcal{S}_{ijkl}((r-1)L^{\zeta / \nu})
\eeq
We find that standard deviation is best collapsed with $\zeta = \zetae \pm \zetaerr$, shown in Fig.~\ref{fig:SigmaCollapse}. The determined value of the exponent $\zeta$ appears to collapse $\sigma L^{1/\nu}$ for all moduli except for $C_{xxxx}$ at larger values of the scaling variable. We also note the nice overlap of data performed at different values of $\left(r,L\right)$ but the same value of the scaling variable $Y$.

We can in principle use information from these collapse plots to make a prediction for the value of the finite-size scaling exponent close to the anisotropic phase transition $\nu^{\textrm{aniso}}$. First, we note that if we fix $r>1$ and send $L\rightarrow\infty$, the spread in the distributions of $p_c$ will narrow as $\sigma\sim L^{-1/\nu^{\textrm{aniso}}}$, as the finite-size effects are (at large enough system sizes) controlled by the critical exponents of the anisotropic critical point. In the scaling function for the distributional spreads of $p_c$, this corresponds to the asymptotic limit $Y\rightarrow\infty$. Forcing the asymptotics of the numerically determined crossover scaling function to agree with the asymptotics expected at the anisotropic critical point will give us a prediction for $\nu^{\textrm{aniso}}$.

Suppose this scaling function has asymptotic behavior $\mathcal{S}_{ijkl}\left(Y\right)\sim Y^{\alpha}$ at large $Y$. Then in the limit $L\rightarrow \infty$ with $r>1$ fixed,
\beq
\sigma L^{1/\nu}\sim Y^{\alpha}\sim L^{\zeta\alpha/\nu}\;\;\textrm{and}\;\;\sigma\sim L^{-1/\nu^{\textrm{aniso}}}
\eeq
together give
\beq
\nu^{\textrm{aniso}}=\frac{\nu}{1-\zeta\alpha},
\eeq
where $\nu$ is the value of the finite-size scaling exponent at the isotropic fixed point. With $\alpha=1.0\pm0.5$, this gives a prediction of $\nu^{\textrm{aniso}}=1.7$, but values between $1.2-3.2$ are consistent with our error bars reported in Table~\ref{table:Exponents}. This is ultimately due to the poor numerical determination of $\zeta$ and $\alpha$. These could also in principle be different for the different $ijkl$; this would be detected through different values of $\alpha$ for each modulus since both $\nu$ and $\zeta$ are properties of the isotropic fixed point.

We estimate the exponent with which each modulus vanishes at their corresponding anisotropic phase transition ($f_{ijkl}^{\textrm{aniso}}$) by considering our largest system size at two values of $r$ away from isotropy ($r=1.2$ and $r=1.5$). The location of the phase transition is determined by averaging the value of $p$ with which each modulus for each lattice becomes rigid.

\begin{figure}[!ht]
    \begin{center}
    \includegraphics[alt={estimate of anisotropic scaling exponent for each modulus}, width=\linewidth]{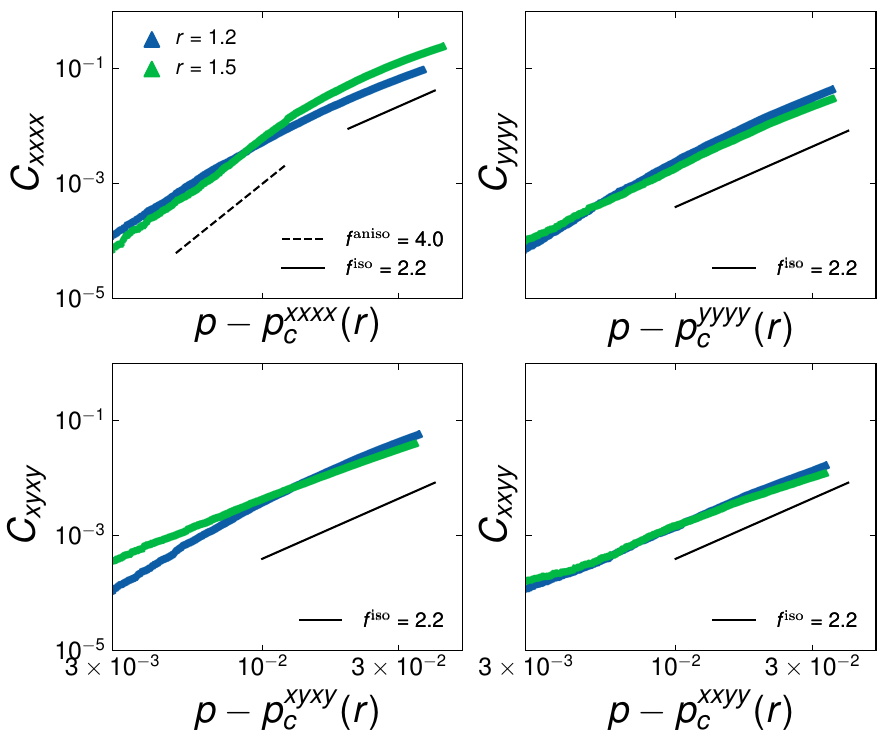}
    \end{center}
    \caption{\textbf{Estimate of anisotropic scaling exponent of each modulus from our largest system size.} The behavior of the $C_{xxxx}$ modulus suggests a crossover between the anisotropic and isotropic scaling exponents, whereas for the other moduli, the behavior appears to be governed by the same isotropic exponent.}
    \label{fig:CrossoverP}
\end{figure}

\begin{figure}[H]
    \begin{center}
    \includegraphics[alt={scaling collapse at constant $Y=0.66$}, width=\linewidth]{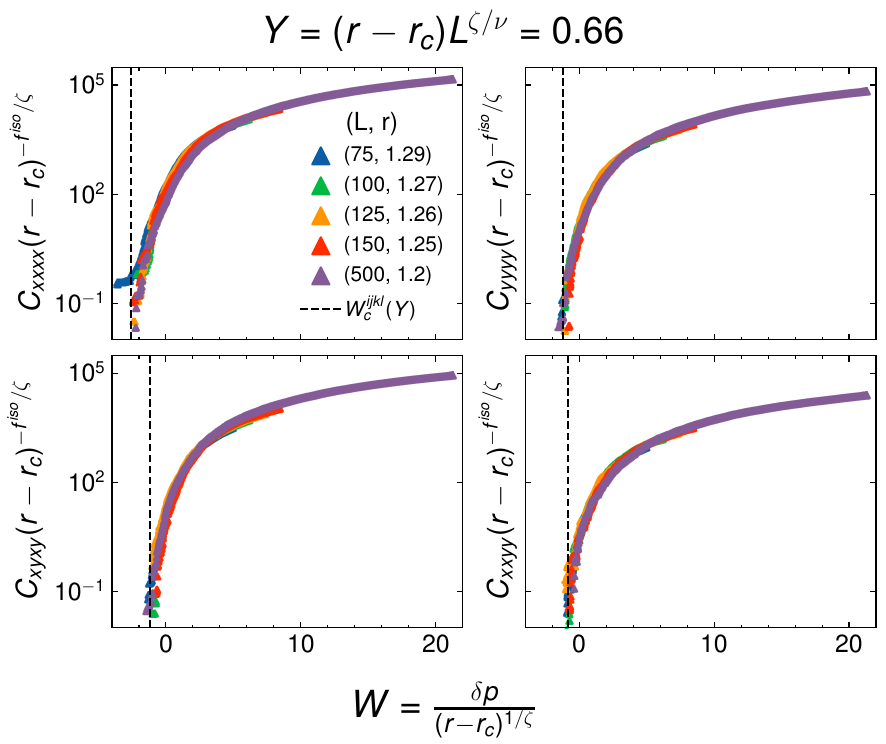}
    \end{center}
    \caption{\textbf{Scaling collapse of all the moduli at constant $Y = 0.66$.} The estimate of $W_c$ is shown with the dashed black line.}
    \label{fig:CollapseLowY}
\end{figure}

\begin{figure}[H]
    \begin{center}
    \includegraphics[alt={scaling collapse at constant $Y=1.65$}, width=\linewidth]{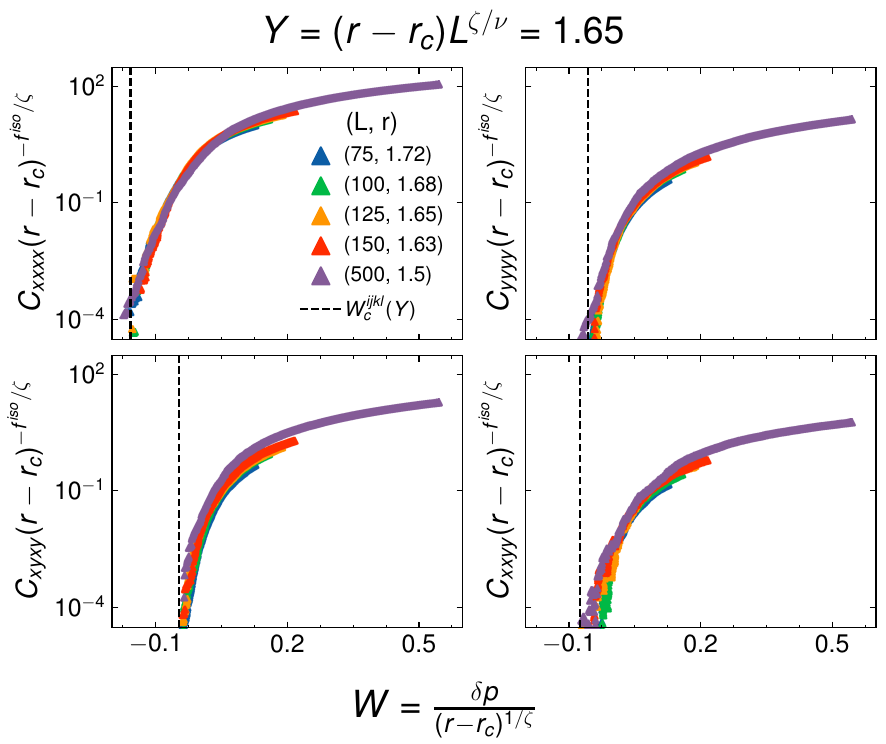}
    \end{center}
    \caption{\textbf{Scaling collapse of all the moduli at a constant $Y = 1.65$.} The estimate of $W_c$ is shown with the dashed black line.}
    \label{fig:CollapseHighY}
\end{figure}

To search for which value of $\zeta$ best collapses the two variable scaling function, we obtain simulations of constant $Y$ across various system sizes. We test a value of $\zeta$ by first fixing our largest system size and then solving for the value of $r$ as a function of system size that results in the same value of $Y$. Figures~\ref{fig:CollapseLowY} and \ref{fig:CollapseHighY} show a scaling collapse of all the moduli with $\zeta=0.25$ and with a constant value of $Y=0.66$ and $Y=1.65$, respectively. The scaling function for each modulus vanishes at a given value of $W$, which we denote as $W_c^{ijkl}(Y)$, as it is dependent on the value of $Y$. We note that further away from isotropy ($Y=0$), there are additional corrections to scaling, resulting in a worse collapse at higher values of $Y$.

\begin{figure}[H]
    \begin{center}
    \includegraphics[alt={estimate of anisotropic scaling exponent for each modulus at constant $Y=0.66$}, width=\linewidth]{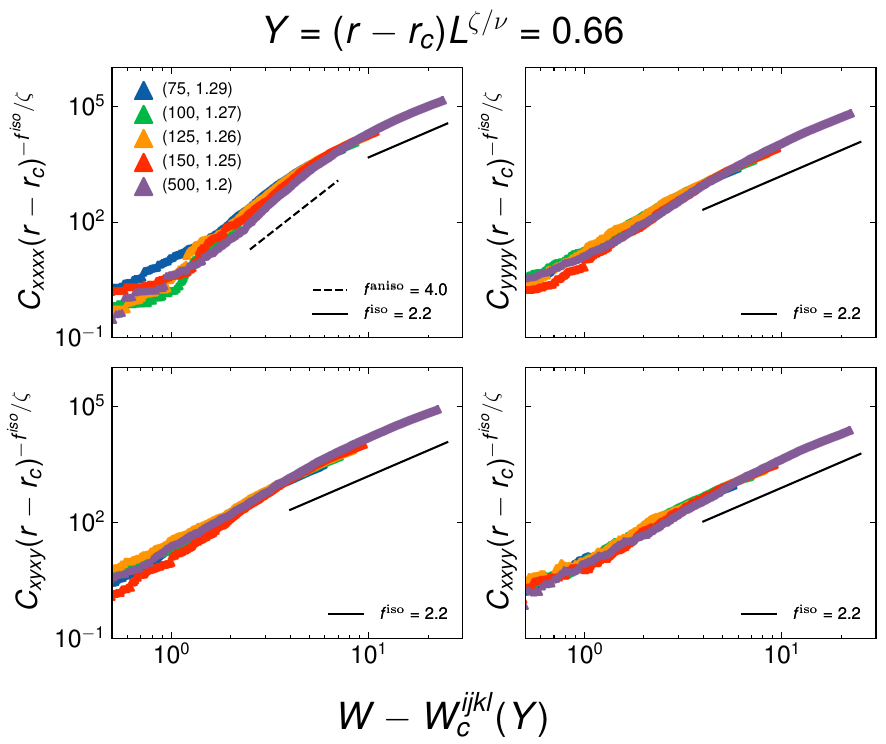}
    \end{center}
    \caption{\textbf{Estimate of anisotropic scaling exponent of each modulus at a constant $Y=0.66$.} The simulation data is the same as found in Fig.~\ref{fig:CollapseLowY}.}
    \label{fig:CrossoverLowY}
\end{figure}

\begin{figure}[H]
    \begin{center}
    \includegraphics[alt={estimate of anisotropic scaling exponent for each modulus at constant $Y=1.65$}, width=\linewidth]{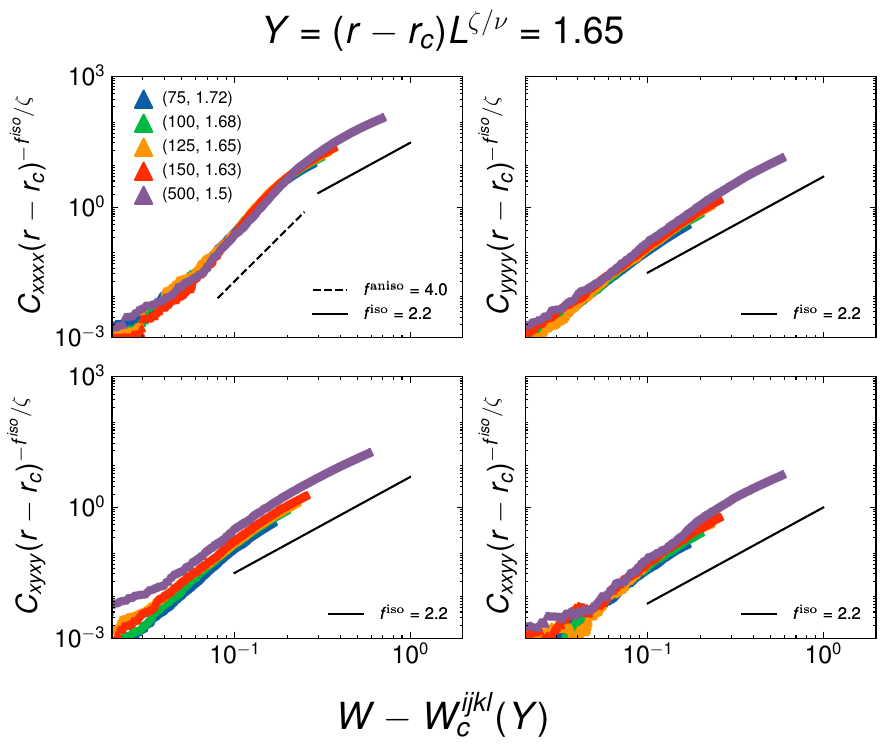}
    \end{center}
    \caption{\textbf{Estimate of anisotropic scaling exponent of each modulus at a constant $Y=1.65$.} The simulation data is the same as found in Fig.~\ref{fig:CollapseHighY}.}
    \label{fig:CrossoverHighY}
\end{figure}

We again obtain an estimate of the scaling exponents in Figures~\ref{fig:CrossoverLowY} and \ref{fig:CrossoverHighY} by plotting the rescaled moduli as a function of distance from $W_c$. The plots suggest crossover for the $C_{xxxx}$ modulus, with $f^\textrm{aniso}_{xxxx}$ governing the behavior for lower values of $W-W_c$ and $f^\textrm{iso}$ for the high $W-W_c$ regime. Furthermore, the other three moduli appear to vanish with a critical exponent indistinguishable from $f^\textrm{iso}$ within our estimated error bars.

\end{document}